\documentclass{article}
\usepackage{spconf,amsmath,amssymb,amsthm,graphicx,url}

\newcommand{\vecr}{\mathbf{r}}
\newcommand{\vecg}{\mathbf{g}}
\newcommand{\vecq}{\mathbf{q}}

\newcommand{\vecx}{\mathbf{x}}
\newcommand{\vecy}{\mathbf{y}}
\newcommand{\vecf}{\mathbf{f}}
\newcommand{\Knoisy}{\widetilde{K}}

\newtheorem{theorem}{Theorem}[section]
\theoremstyle{definition}

\title{Gaussian process regression can turn non-uniform and undersampled diffusion MRI data into diffusion spectrum imaging}
 \name{Jens Sj\"{o}lund$^{\star \dagger \ddagger}$ \qquad Anders Eklund$^{\star \dagger \S}$ \qquad Evren \"{O}zarslan$^{\star}$ \qquad Hans Knutsson$^{\star \dagger}$\thanks{Research supported by the Swedish Research Council (VR) grants 2012-4281, 2013-5229 and 2015-05356, the Swedish Foundation for Strategic Research (SSF) grant AM13-0090, EUREKA ITEA BENEFIT grant 2014-00593 and the Linneaus center
CADICS. Data collection and sharing for this project was provided by the Human Connectome Project (HCP; Principal Investigators: B. Rosen, A. W. Toga, V. J. Weeden). HCP funding was provided by NIDCR, NIMH and NINDS.}}

 \address{$^{\star}$ Department of Biomedical Engineering, Link\"{o}ping University, Link\"{o}ping, Sweden \\
     $^{\dagger}$ Center for Medical Image Science and Visualization,  Link\"{o}ping University, Link\"{o}ping, Sweden\\
$^\ddagger$ Elekta Instrument AB, Kungstensgatan 18, Box 7593, SE-103 93 Stockholm, Sweden\\
$^\S$ Department of Computer and Information Science, Link\"{o}ping University, Link\"{o}ping, Sweden}
\begin{document}
%\ninept
%
\maketitle
\begin{abstract}
We propose to use Gaussian process regression to accurately estimate the diffusion MRI signal at arbitrary locations in q-space. By estimating the signal on a grid, we can do synthetic diffusion spectrum imaging: reconstructing the ensemble averaged propagator (EAP) by an inverse Fourier transform. We also propose an alternative reconstruction method guaranteeing a nonnegative EAP that integrates to unity. The reconstruction is validated on data simulated from two Gaussians at various crossing angles. Moreover, we demonstrate on non-uniformly sampled in vivo data that the method is far superior to linear interpolation, and allows a drastic undersampling of the data with only a minor loss of accuracy. We envision the method as a potential replacement for standard diffusion spectrum imaging, in particular when acquistion time is limited.
\end{abstract}
\section{Introduction}
\label{sec:intro}

%\subsubsection{General intro to dMRI}
The structure of biological tissue affects the diffusion of particles within it. The diffusion MRI signal stems from the translational motion of spin-bearing particles---water in particular---and can thus be used to probe the tissue structure. 

% Ensemble averaged propagator
The voxel-averaged distribution of spins' displacement is referred to as the ensemble averaged propagator (EAP). Although the imaging voxel is macroscopic, the characteristic length-scale of the translational motion in a typical diffusion MRI experiment is commensurate with cell dimensions. This makes the EAP an indirect but potentially powerful way of describing the diffusion behavior in complex materials \cite{Tuch2002}.
%\subsection{Applications}
For example, the EAP can be used to characterize the tissue using various scalar indices \cite{Tuch2002, Ozarslan2013, Ning2015}. It can also be used in tractography by mapping the EAP to an orientation density function (ODF) through a radial projection \cite{Wedeen2008}. This enables the whole arsenal of tractography algorithms developed for High Angular Resolution Diffusion Imaging (HARDI) \cite{Descoteaux2015}.

Under the narrow pulse approximation \cite{Stejskal1965}, there is a direct Fourier relationship \cite{Stejskal1965b} between the normalized diffusion signal, $E(\vecq) = S(\vecq)/S(0)$, in q-space and the EAP, denoted $P(\vecr)$, in real space 
\begin{equation}
P(\vecr) = \frac{1}{(2\pi)^3}\int_{\vecq\in\mathbb{R}^3} E(\vecq)\exp(i\vecq\cdot\vecr)d\vecq,\label{eq:FourierRelationship}
\end{equation}
where $\vecr$ is the displacement vector and $S(\vecq)$ is the diffusion signal measured at q-space point $\vecq=\gamma\delta\vecg$. Here, $\gamma$ is the gyromagnetic ratio and $\delta$ is the duration of the diffusion sensitizing gradients whose magnitude and orientation are determined by the vector $\vecg$. 
%In the language of probability theory, the normalized signal is the characteristic function of the averaged propagator.
%The fundamental relation \eqref{eq:FourierRelationship} implies that $P(\vecr)$ is uniquely determined by the $E(\vecq)$ which is measured directly in experiments. Thus, the propagator estimation problem reduces to one of estimaing a continuous function $E(\vecq)$ based on scattered measurements acquired in q-space.
%\subsubsection{Diffusion spectrum imaging}
Diffusion Spectrum Imaging (DSI) \cite{Wedeen2005} is the direct application of equation \eqref{eq:FourierRelationship}: measurements are done on a dense Cartesian grid in q-space, after which the EAP is found by an inverse 3D Fourier transform.  However, this requires such a large number of q-space samples that the acquisition time becomes too long for routine use. 

We describe how to use a machine learning method called Gaussian process regression to estimate the q-space signal based on far fewer samples than required by DSI. In particular, we show how to resample data acquired on multiple shells (radii) onto a Cartesian grid. Although an estimate of the EAP could then be obtained by an inverse Fast Fourier Transform, we also describe a theoretically well-founded reconstruction method that respects the probabilistic nature of the EAP.

%\subsubsection{Related work}
%Considering the potential range of applications and the shortcomings of DSI, it is perhaps unsurprising that 
Other researchers have also tackled this problem. One approach is to expand the q-space signal in a suitably chosen basis, e.g. in Hermite functions \cite{Ozarslan2013} or using a mixture of radial basis functions densely distributed in q-space \cite{Ning2015}. %Both of these have the advantage that they enable an analytic Fourier inversion.
Although both of these can in theory approximate any function to arbitrary accuracy, the finite number of samples in combination with the ever present noise, especially at large q-values, places severe restrictions on these models. 
%The Hermite function expansion has to be truncated at a fairly low order (typically $n=6$), which limits is flexibility. The expansion in radial basis functions, while flexible, requires the fitting of a large number of parameters (possibly more than the number of samples) which is always a risky endeavour, in particular if the samples are few and far between. 
Other methods require a particular sampling scheme, e.g. multiple concentric shells, after which some type of interpolation \cite{Wu2007} or smoothing \cite{Descoteaux2011} can be used to estimate the intermediate values.  

Adopting a Gaussian process framework, such as the one we propose, immediately gives access to an extensive set of tools that unite an elegant probabilistic view with computational tractability \cite{Rasmussen2006}. Notably, it enables expressive models yet has few parameters and it comes with a rigorous way of reasoning about uncertainty \cite{Wilson2014}. This work was inspired by previous work on using Gaussian processes to correct for artifacts in diffusion scans \cite{Andersson2015}. 

\section{Theory}

\subsection{Gaussian process regression}
A Gaussian process can be thought of as a Gaussian distribution over functions \cite{Rasmussen2006}.  Just as a multivariate Gaussian distribution is fully specified by its mean and covariance matrix, a Gaussian process is fully described by its mean and covariance function. 
%To make this concrete, consider a stochastic variable $y$ distributed according to a multivariate normal distribution with mean $\mu$ and covariance matrix $\Sigma$
%\begin{equation}
%y \sim \mathcal{N}(\mu, \Sigma).
%\end{equation}
%The covariance between two points $x_i$ and $x_j$ are specified by the entry $\Sigma_{ij}$ in the covariance matrix, or if you imagine $\Sigma$ being a function, you could write this as $\Sigma_{ij} = \Sigma (x_i, x_j) $.
%
%To make the transition to a Gaussian process, we note that a function $f$ can be loosely be thought of as a very long vector, where each entry in the vector $f(x)$ corresponds to the function value at the input point $x$. Imagine taking $N$ uniformly spaced input points $x_i$, $i = 1,\ldots, N$, and suppose that $y = f(x) \sim \mathcal{N}(\mu,\Sigma)$ where, as usual, the mean and covariance are defined by 
%\begin{align}
%\mu &= E[y] \in \mathbb{R}^N,\\
%\Sigma&= E[(y-\mu)(y-\mu)^T]\in \mathbb{R}^{N\times N}.
%\end{align}
%In the limit $N\to\infty$, we get a Gaussian \emph{process} 
%\begin{equation}
%f(x) \sim \mathcal{GP}(\mu (x), k(x,x')),
%\end{equation}
%where $m(x)$ is the mean function and $k(x, x')$ the covariance function (analogous to an entry in the covariance matrix $\Sigma (x, x')$). In contrast to the finite dimensional case, the mean and covariance functions are defined for every value of $x$ and $x'$. 
However, the mean function is typically set to zero because its effects can be absorbed in the covariance function. The covariance function tells us how similar two inputs are and thus how much they are allowed to influence each other. 
%The choice of covariance function therefore controls the type of solution we expect, e.g. if it is smooth, periodic or has some particular symmetry.
The evaluation $k(x_i,x_j) $ of the covariance function at two points $x_i$ and $x_j$ is akin to the entry $\Sigma_{ij}$ of a covariance matrix $\Sigma$.
%We can thus appreciate that for $k(x, x')$ to be a valid covariance function, any matrix $K$ with elements $k(x_i, x_j)$ must be symmetric and positive definite, just like a covariance matrix.  
%Incidentally, this allows us to glean one of the beauties of Gaussian processes: although it involves infinite dimensional objects, it reverts back to ``regular'' multivariate Gaussian distributions as long as you consider the function values only at a finite set of points.

%\subsubsection{Gaussian process regression}
Now, let us describe how to use Gaussian processes for regressions provided that the mean and covariance functions are known.
Suppose we have a training set $\mathcal{D}= \{(\vecx_i, y_i)|\,i=1,\ldots, n\}$ composed of inputs $\vecx_i\in\mathbb{R}^m$ and noisy measurement values $y_i\in\mathbb{R}^1$ generated from a latent function $f(\vecx)$ with i.i.d. Gaussian noise such that $y_i=f(\vecx_i)+\epsilon_i$.
Gaussian processes allow you to predict the function value $f_*=f(\vecx_*)$ at an arbitrary input point $\vecx_*$, as well as the corresponding variance $\sigma_*^2$ (uncertainty if you will). If we organize the inputs as a matrix $X$ with rows $\vecx_i$ and the function values as a vector $\vecf$, the joint distribution of the training data and the unobserved pair $(\vecx_*, f_*)$ is given by
\begin{equation}
\begin{pmatrix}
\vecf\\
f_*
\end{pmatrix}
\sim
\mathcal{N}_{n+1}\left(
\begin{pmatrix}
\boldsymbol{\mu(\vecx)}\\
\mu(x_*)
\end{pmatrix},
\begin{pmatrix}
K+\sigma_n^2 I&\mathbf{k}_*\\
\mathbf{k}_*^T &k_*
\end{pmatrix}
\right),
\end{equation}
where $K$ is the $n\times n$ matrix of covariances between all points in the training data, $\sigma_n^2$ is the noise variance, $\mathbf{k}_*$ is an $n\times 1$ vector of cross-covariances between the training data and the unobserved point $x_*$. Finally, $k_* = k(x_*,x_*)$  is the variance at the point $x_*$. We use the convention that upper case signifies a matrix and bold font signifies a vector. 

Using the standard identity for conditioning a Gaussian distribution \cite{Rasmussen2006}, we find that the conditional probability of $f_*$ given $X$, $\vecf$ and $x_*$ is
\begin{align}
p(f_*|X, \vecf, x_*) &= \mathcal{N}(\mu_*, \sigma_*^2), \quad \text{where}\\
\mu_* &= \mu(x_*) + \mathbf{k}_*^T (K+\sigma^2 I)^{-1}\vecy,\label{eq:predictiveMeanWithNoise}\\
 \sigma_*^2 &= k_*-\mathbf{k}_*^T(K+\sigma^2 I)^{-1}\mathbf{k}_*\label{eq:predictedVarianceWithNoise}. 
\end{align}
%The most probable value of $f_*$ is thus $f_* = \mu_*$, i.e. the mean after conditioning on the observed values. 

\subsection{A covariance function for diffusion MRI}
There are several characteristics of the diffusion MRI signal that we would like to capture by an appropriate choice of covariance function.
First, the signal is expected to be symmetric about the origin, i.e. $E(-\vecq) = E(\vecq)$. Second, we do not expect there to be any preferential directionality in the sense that the covariance between the signal when measuring in two directions should only depend on the angle between them. This suggests a factorization of the covariance into a radial part $C_r$ and an angular part $C_\theta$ such that 
\begin{equation}
k(\vecq_i, \vecq_j) = C_r(q_i, q_j)C_\theta (\hat{\vecq}_i \cdot \hat{\vecq}_j)
\end{equation}
where $q = |\vecq|$ and $\hat{\vecq}=\vecq/|\vecq|$. In order for the angular part to be a valid covariance function on the sphere, we use the following theorem \cite{Schoenberg1942, Huang2011}
\begin{theorem}
A real continuous function $C(\theta)$ is a valid covariance function on the sphere if and only if it is of the form
\begin{equation}
C(\theta) = \sum_{n=0}^{\infty} a_n P_n(\cos \theta),\quad \theta \in [0,\pi] \label{eq:angularTheorem}
\end{equation}
where $a_n\geq 0$, $\sum_{n=0}^\infty a_n < \infty$ , and $P_n(\cdot)$ are the Legendre polynomials.
\end{theorem}
We take $C_\theta$ equal to the even terms of order $n\leq 6$ in the above sum. Excluding odd terms guarantees symmetry about the origin, as desired.

We parameterized the radial part as
\begin{equation}
C_r(q_i, q_j) = \exp\left(-\frac{1}{2\sigma_r^2}\left(\log\left(\frac{\xi^2+q_i^2}{\xi^2+q_j^2}\right)\right)^2\right),\label{eq:radialCov}
\end{equation}
where $\xi \ll q$ is a constant used to make the function continuous in the origin.
This is a valid kernel since it is the composition of a radial basis function with a function $\psi(q) = \log(\xi^2+q^2)$ \cite{Wilson2014}.

%\subsubsection{Training a GP --- Learning the hyperparameters}
Taken together, we end up with six hyperparameters $\boldsymbol \eta$: four coefficients for the angular covariance \eqref{eq:angularTheorem}, a length-scale $\sigma_r$ of the radial covariance \eqref{eq:radialCov} and the noise variance $\sigma_n^2$. These are estimated by maximizing the marginal likelihood (so called empirical Bayes or type II maximum likelihood). Up to a constant, the logarithm of the marginal likelihood is given by \cite{Rasmussen2006}
\begin{equation}
\log p(\vecy | \vecx, \boldsymbol{\eta}) = -\frac{1}{2}\vecy^T \Knoisy^{-1} \vecy - \frac{1}{2}\log |\Knoisy|,
\end{equation}
where $\Knoisy = K + \sigma_n^2 I$ is the covariance matrix of the noisy measurements. If the voxels are assumed to be independent, then $\Knoisy$ is block-diagonal: $\Knoisy = I \otimes \Knoisy_\text{voxel}$, where $\otimes$ denotes the Kronecker product. This means that the marginal likelihood factorizes as
\begin{equation}
\log p(\vecy | X, \boldsymbol{\eta}) = -\frac{1}{2}\sum_{v\in V} \left(\vecy_v^T \Knoisy_\text{voxel}^{-1}\, \vecy_v +\log |\Knoisy_\text{voxel}|\right) ,
\end{equation}
where $V$ is the set of all voxels. In practice, we perform the hyperparameter estimation on a large subset of all voxels and the testing on a different subset. Our implementation was made in MATLAB \cite{MATLAB2016a} using the Gaussian Processes for Machine Learning (GPML) toolbox \cite{Rasmussen2010}. 

\subsection{Reconstruction of the ensemble averaged propagator}
Recall from equation \eqref{eq:FourierRelationship} that the ensemble averaged propagator (EAP) is the inverse Fourier transform of the normalized signal. A simple method for computing the EAP involves interpolating the signal onto a Cartesian grid and then applying a fast Fourier transform (FFT). However, this procedure does not guarantee that the resulting EAP estimate $\hat{P}(R)$ is nonnegative and integrates to unity. The quick and dirty solution for this is to set negative values to zero and then renormalize. We will however consider a better founded approach. In short, we readjust the estimated signal, using the variances of the Gaussian process estimates as weights, such that the inverse Fourier transform is nonnegative and integrates to unity. 

The Gaussian process estimate at $\vecx_*$ is $f_* \sim \mathcal{N}(\mu_*,\sigma_*^2)$ where the mean and variance are given by equations \eqref{eq:predictiveMeanWithNoise} and \eqref{eq:predictedVarianceWithNoise} respectively. Since the predictions for different inputs are conditionally independent, the resulting log-likelihood is 
\begin{align}
\log p(\vecf_*|\boldsymbol{\mu_*},\Sigma_*) &= -\frac{1}{2}(\vecf_*-\boldsymbol{\mu}_*)^T \Sigma_*^{-1}(\vecf_*-\boldsymbol{\mu}_*)\nonumber\\
&= -\frac{1}{2}\|W(\vecf_*-\boldsymbol{\mu}_*)\|_2^2,\label{eq:weightedLS}
\end{align}
 where we have introduced a weight matrix $W = \text{diag}(\sigma_{*}^{-1})$.

The discrete inverse Fourier transform can be expressed as a matrix, which we denote $F$. The constraint on nonnegative probability estimates can then be written simply as $F\vecf_*\geq 0$. To integrate to unity, it must hold that $f_* = 1$ when $\vecx_*=0$. The nature of the diffusion signal requires it to be nonnegative, this is included as a bound. We thus end up with the following constrained weighted least-squares problem:
\begin{equation}
\begin{aligned}
& \underset{\vecf_*}{\text{minimize}}
& & \|W(\vecf_*-\boldsymbol{\mu})\|_2^2 \\
& \text{subject to} & & F\vecf_* \geq 0\\
&&& \vecf(0)=1\\
&&& \vecf_*\geq 0.
\end{aligned}
\end{equation}
This is a convex quadratic programming problem which can be efficiently solved to global optimiality using e.g. an interior-point method.

\subsection{Data augmentation}
The Gaussian process model typically excels at interpolation and smoothing, whereas extrapolation poses more difficulties. To improve the extrapolation ability, we augment the data set with synthetic data at the origin (signal equal to one) and at a large radius where the signal is set to zero. Outside this cut-off radius, all signal estimates are set to zero. The data augmentation is done after training, so as to not affect the hyperparameters learned, but before prediction.

\section{Results}
\subsection{Simulated data}
\begin{figure}[htb]
\begin{minipage}[b]{.48\linewidth}
  \centering
  \centerline{\includegraphics[width=4.0cm]{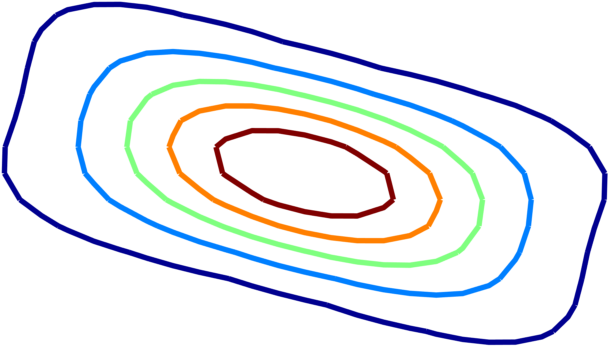}}
%  \vspace{1.5cm}
  \centerline{(a) Exact, $\phi=30^\circ$}\medskip
\end{minipage}
\hfill
\begin{minipage}[b]{0.48\linewidth}
  \centering
  \centerline{\includegraphics[width=4.0cm]{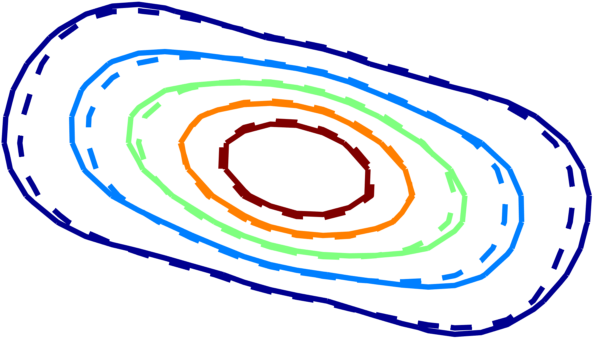}}
%  \vspace{1.5cm}
  \centerline{(b) Estimated, $\phi=30^\circ$ }\medskip
\end{minipage}
\begin{minipage}[b]{.48\linewidth}
  \centering
  \centerline{\includegraphics[width=4.0cm]{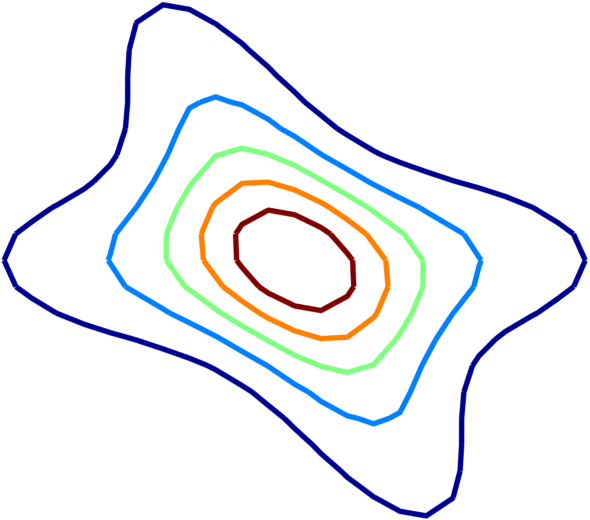}}
%  \vspace{1.5cm}
  \centerline{(a) Exact, $\phi=60^\circ$}\medskip
\end{minipage}
\hfill
\begin{minipage}[b]{0.48\linewidth}
  \centering
  \centerline{\includegraphics[width=4.0cm]{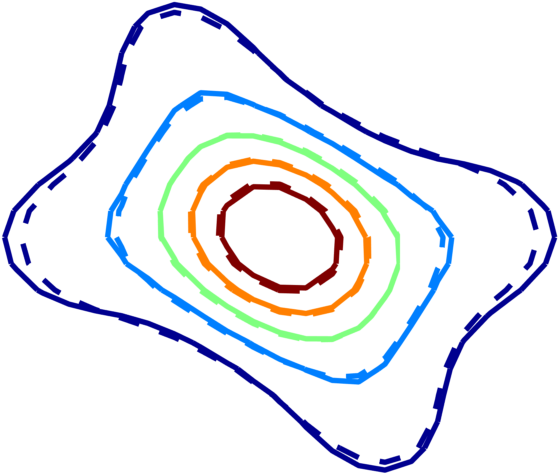}}
%  \vspace{1.5cm}
  \centerline{(b) Estimated, $\phi=60^\circ$ }\medskip
\end{minipage}
\begin{minipage}[b]{.48\linewidth}
  \centering
  \centerline{\includegraphics[width=4.0cm]{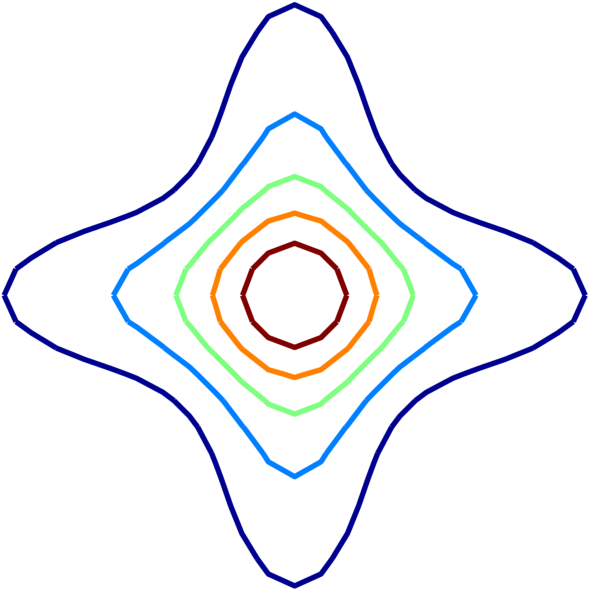}}
%  \vspace{1.5cm}
  \centerline{(a) Exact, $\phi=90^\circ$}\medskip
\end{minipage}
\hfill
\begin{minipage}[b]{0.48\linewidth}
  \centering
  \centerline{\includegraphics[width=4.0cm]{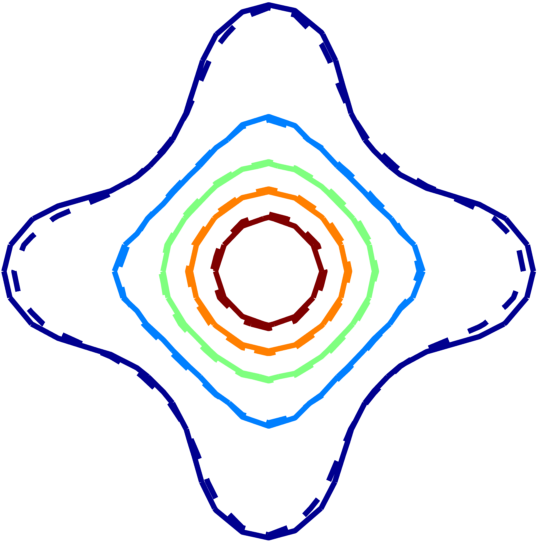}}
%  \vspace{1.5cm}
  \centerline{(b) Estimated, $\phi=90^\circ$ }\medskip
\end{minipage}
\caption{Exact (left) and reconstructed (right) averaged propagators for data simulated from two equal magnitude Gaussians at various crossing angles $\phi$. Contour plots of the x-y plane are shown with equidistant level sets. Dashed and continuous lines indicate linear interpolation and our method, respectively.}
\label{fig:contourPlots}
\end{figure}

We simulated data from two Gaussians of equal magnitude and equal but rotated diffusion tensors. As the unrotated diffusion tensor, we used $D_1 = D_0\cdot\text{diag}(1, 0.1, 0.1)$, where $D_0 = 2.5\cdot 10^{-9}$ m$^2$/s. This yields a mean diffusivity (MD) of 1 $\mu$m$^2$/ms and a fractional anisotropy (FA) of 0.89, which is roughly the value observed in the white matter of the brain. The second diffusion tensor was determined by rotating $D_1$ by an angle $\phi$ about the z-axis, such that $D_2 = R(\phi)^T D_1 R(\phi) $. The latent signal was thus
\begin{equation}
E(\vecq, \phi) = \frac{1}{2}\left(e^{-t_d\, \vecq^T D_1 \vecq} + e^{-t_d\, \vecq^T D_2 \vecq}\right),
\end{equation}
where $t_d = \Delta - \delta/3$; here $\Delta$ is the mixing time and $\delta$ the pulse duration. The latent signal was corrupted with Rician noise with scale parameter $\sigma = 0.01$ to yield the simulated signal.
%and $\epsilon$ denotes i.i.d. Gaussian noise with standard deviation 0.01.
%\begin{equation}
%P(\vecr, \Delta) = \sum_{i=1}^N w_i \cdot \mathcal{N}(0,2\Delta \cdot D_i)
%\label{eq:averagedPropagatorTwoGaussians}
%\end{equation}
%%For clarity, we restrict attention to two Gaussians of equal weights, $w_1=w_2=0.5$, and equal but rotated diffusion tensors: $D_1 = D_0\cdot\text{diag}(1, 0.1, 0.1)$ and $D_2 = R(\theta)^T D_1 R(\theta)$, where $D_0 = 2.5\cdot 10^{-9}$ m$^2$/s. These diffusion tensors have a mean diffusivity (MD) of 1 $\mu$m$^2$/ms and fractional anisotropy (FA) of 0.89, which is roughly the same as in the white matter of the brain. 
%The signal  which is 
%\begin{equation}
%E(\vecq, \Delta) = \sum_{i=1}^N w_i\cdot \exp\left(-\Delta \vecq^T D_i \vecq\right).
%\end{equation}
%We take this as the simulated signal and add iid Gaussian noise with standard deviation 0.05 to form the simulated measurement. This noise level is of the same order of magnitude as the HCP data. 
We used the same experimental parameters as in the the Human Connectome data described in the next section. The hyperparameters were optimized on a set of 100 Gaussian mixtures with randomly sampled crossing angles. Figure \ref{fig:contourPlots} compares exact and reconstructed EAPs for $\phi = 30^\circ, 60^\circ, 90^\circ$. We compare with using linear interpolation as in Hybrid Diffusion Imaging (HYDI) \cite{Wu2007}. Table \ref{table:RTOP} shows the relative error in the estimation of the return-to-origin probability, $P(0)$, which is a scalar index indicative of the underlying structure.

\begin{table}
\centering
{\renewcommand{\arraystretch}{1} 
\begin{tabular}{c|cc}
%&\multicolumn{2}{c}{$|P(0)-\hat{P}(0)|/P(0)$}\\%
%\cline{2-3}
 $\phi$&This work&Linear interp.\\
\hline
\rule{0pt}{2.5ex} $30^\circ$&0.036&0.058\\
$60^\circ$&0.030&0.051\\
$90^\circ$&0.027&0.050
\end{tabular}
}
\caption{Relative error in estimation of return-to-origin probability $P(0)$ for different crossing angles $\phi$.}
\label{table:RTOP}
\end{table}

\subsection{Reconstruction of subsampled in vivo data}
We used in vivo diffusion data obtained from the Human Connectome Project (HCP) \cite{vanEssen2013}
database\footnote{\url{http://www.humanconnectome.org/documentation/MGH-diffusion/}}. The subjects are healthy adults, scanned on a customized Siemens 3T Connectom scanner \cite{Setsompop2013, Keil2013} using a Stejskal-Tanner type diffusion weighted spin-echo sequence. Diffusion measurements were at four b-value shells ($b = t_d q^2$): 1000, 3000, 5000, 10000 s/mm$^2$. The corresponding number of gradient orientations were 64, 64, 128 and 256. %A total of 40 non-diffusion weighted (b$=0$) images was acquired at regular intervals. 

\begin{figure}[htb]
\begin{minipage}[b]{1.0\linewidth}
  \centering
  \centerline{\includegraphics[width=8.5cm]{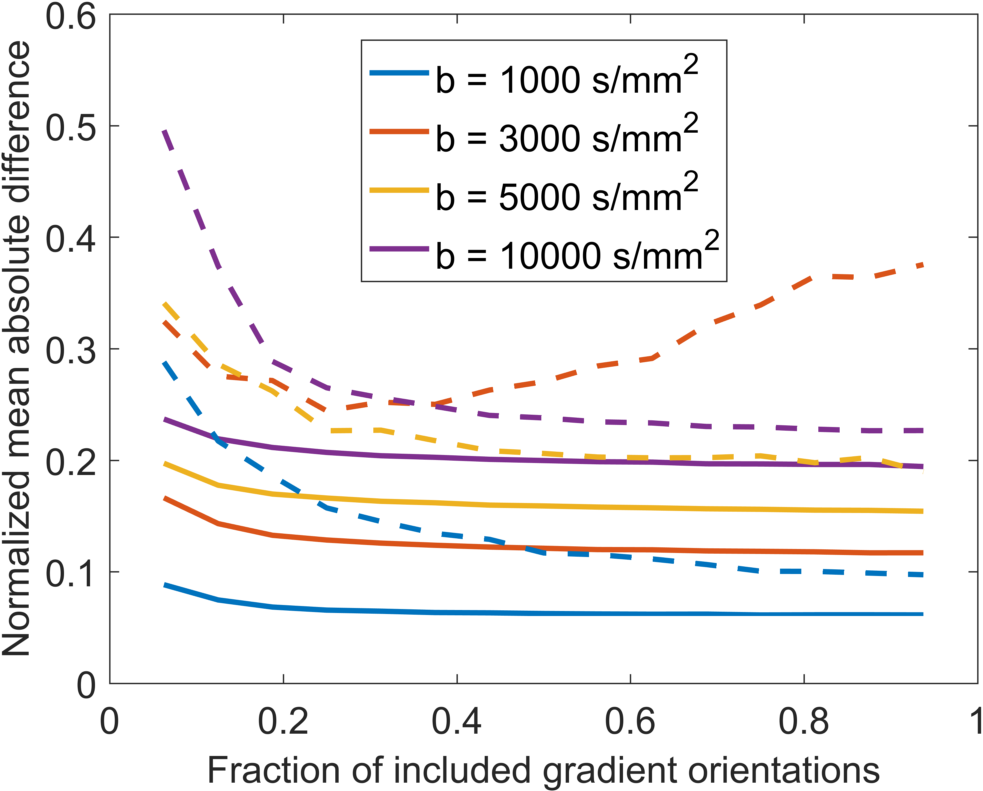}}
%  \vspace{2.0cm}
  %\centerline{(a) Result 1}\medskip
\end{minipage}
\caption{Mean absolute differences, scaled by the respective global shell means, as measurements are removed and instead estimated. Equal fractions of gradient orientations where randomly removed from each shell. Dashed and continuous lines indicate linear interpolation and our method, respectively.}
\label{fig:subsamplingHCP}
\end{figure}

To illustrate that the proposed method performs well even as the data is severely undersampled, we randomly exclude an equal fraction of measurements from each shell and instead estimate the signal value. To compensate for statistical fluctuations due to the sampling, we averaged the errors over 10 realizations. The hyperparameters were optimized on a set of 100,000 voxels from the same subject. Figure \ref{fig:subsamplingHCP} shows the average differences between measurements and estimates as computed on another set of 10,000 voxels from the same subject. Also here, we compare with linear interpolation \cite{Wu2007}.

\section{Discussion and conclusions}
From figure \ref{fig:subsamplingHCP}, it is clear that our method is superior to linear interpolation and performs well even when the data is drastically undersampled, e.g. 20\% of the gradient orientations gives comparable performance as when having 95 \%. %Incidentally, our method is also faster than scattered linear interpolation (not shown).

We expected the errors in figure \ref{fig:subsamplingHCP} to decay monotonically to a constant, noise-dependent, value. 
So, the poor performance of the interpolation at $b = 3000$ s/mm$^2$  warrants an explanation. A closer inspection (not shown) reveals that the interpolation consistently overestimates the signal in this shell. This is due to HCP's sampling pattern (same gradient orientations used in multiple shells), which  leads to a predominantly radial interpolation. However, the signal decay is convex in this range, so linear interpolation yields an overestimation of the signal.
It is likely that other sampling schemes could alleviate this problem \cite{Ye2012, Knutsson2013}. The same could also be said if the aim is to reconstruct the orientation distribution function (ODF) instead of the EAP. 

Figure \ref{fig:contourPlots} shows that the reconstructed EAPs are similar to the exact EAP, albeit somewhat smoother. Qualitatively, the EAPs reconstructed using our method and linear interpolation appear very similar, but table \ref{table:RTOP} shows that our method gives a considerably more accurate estimate of the return-to-origin probability. We hypothesize that even better reconstructions would be achievable if the sampling pattern was optimized considering the inherent covariance structure of the signal.

%Other choices for mean function, e.g. MAP-MRI? Common practice is to assume a constant mean and estimate it as the empirical mean of the data. However, as pointed out in \cite{Wilson2014}, the mean function can be used as a powerful way to encode assumptions into a Gaussian process model.\\

%Here we used empirical Bayes to train, would be possible to go Bayesian all the way by specifying prior distributions for the parameters.\\

For computational efficiency, we assumed that voxels can be treated independently. It is, however, straightforward to encode spatial dependence through the covariance function. %This is another topic for future research.

%You can get an analytical representation of the EAP as the inverse Fourier transform of the GP estimate if you like\\

%noise model: actually Rician, but this is  approximately Gaussian for SNR $\geq$ 3 (cite). The correct way would be to use the "breaking the noise floor" - method of Koay et al. that converts Rician noise to Gaussian. 

In conclusion, we have used a Gaussian process framework to estimate the diffusion MRI signal and reconstruct the EAP. We have demonstrated the efficacy of the estimation on non-uniform, drastically undersampled in vivo data. We envision the method as a potential replacement for standard diffusion spectrum imaging when acquistion time is limited.

\vfill
\pagebreak

% References should be produced using the bibtex program from suitable
% BiBTeX files (here: strings, refs, manuals). The IEEEbib.bst bibliography
% style file from IEEE produces unsorted bibliography list.
% -------------------------------------------------------------------------
\bibliographystyle{IEEEbib}
\bibliography{strings,refs}

\end{document}